\documentclass[preprint,12pt]{elsarticle}

\usepackage{amssymb}
\usepackage{amsmath}
\usepackage{makecell}

\journal{Optics and Laser Technology}

\begin{document}

\begin{frontmatter}



\title{Analysis of multi-pass pumped thin-disk laser performance with measured disk deformation}

\author[label1]{Hanjin Jo\corref{cor1}}
\ead{hanjin.jo@hilase.cz}
\cortext[cor1]{Corresponding author}
\author[label1]{Jiří Mužík}
\author[label1]{Pawel Sikocinski}
\author[label1]{Magdalena Sawicka-Chyla}
\author[label1]{Michal Chyla}
\author[label1]{Yuya Koshiba}
\author[label1]{Yoann Levy}
\author[label1]{Kohei Hashimoto}
\author[label1]{Martin Smrž}
\author[label1]{Tomáš Mocek}

 \affiliation[label1]{organization={HiLASE Centre, Institute of Physics of the Czech Academy of Sciences},
             city={Za Radnicí 828, Dolní Břežany},
             postcode={252 41},
             country={Czech Republic}
             }

\begin{abstract}
Predicting the steady-state performance of high-power thin-disk lasers requires not only pump-signal energy transfer but also how disk deformation contributes intra-cavity mode formation. In this work, we address the output-power reduction that occurs even when the laser remains in a single-mode regime with $\mathrm{M^2}$ around 1.1. We developed a numerical model in which the pump-induced inversion is initialized from a non-lasing multi-pass absorption model and then coupled to two-dimensional cavity-field propagation using measured disk optical path difference (OPD) maps. Applied to the Yb:YAG thin-disk laser, the model reproduces the residual pump fraction and predicts the signal power, beam diameter, and $\mathrm{M^2}$ with errors of 3.0$\%$, 1.7$\%$, and 0.05, respectively. To interpret the measured OPD, the disk surface is further analyzed by Zernike decomposition, and the defocus term is converted into an equivalent radius of curvature (eROC). The eROC-based simulation provides the defocus-only reference performance that is theoretically reachable in the absence of higher-order aberrations, whereas the measured-OPD simulation reproduces the experimentally observed power reduction at high pump intensity. The comparison quantitatively shows that higher-order aberrations beyond defocus reduce the overlap with the fundamental cavity mode and limit power scaling, even before strong $\mathrm{M^2}$ degradation appears. This result identifies aberration-induced modal loss as a key limitation in high-power single-mode thin-disk lasers.
\end{abstract}

\begin{graphicalabstract}
\centering
\includegraphics[width=11cm]{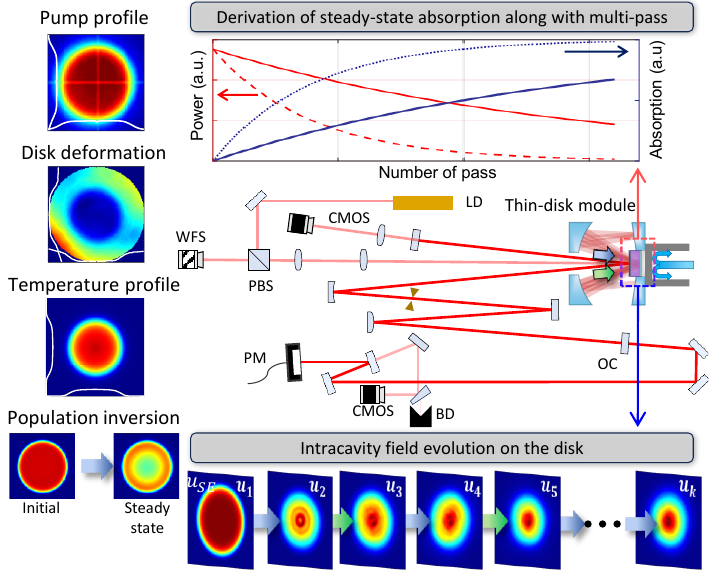}
\end{graphicalabstract}

\begin{highlights}
\item Non-lasing pump initialization enables efficient 2D cavity modeling.
\item Measured OPD maps reproduce single-mode thin-disk laser performance.
\item Curvature-only modeling defines a theoretically achievable output.
\item Higher-order disk aberrations hinders fundamental-mode formation.

\end{highlights}


\end{frontmatter}



\section{Introduction}
\label{Introduction1}
The thin-disk (TD) laser is an attractive gain geometry for achieving excellent beam quality and high signal power owing to relatively low-cost and superior thermal management. Accordingly, TD lasers have been used over a broad spectral range from the mid-IR to the DUV in continuous-wave (CW), micro- to nanosecond pulse, and ultrafast regimes~\cite{MIDIR,MIDIR2,MIDIR3,nsLaser,fsLaser}. 

Despite the favorable thermal management of TD, the signal power does not scale linearly with pump power~\cite{Scale1}. As the pump intensity increases, the difference between the theoretically expected signal power predicted from energetics point of view and that measured experimentally gradually diverges. In thin-disk structures, disk properties such as the absorption distribution, temperature field, thermal properties, refractive-index change, pump–signal overlap, and surface deformation directly affect resonator behavior~\cite{TL1,Las2,BS1,AMP1,AMP2}. Therefore, the performance limitations of high-power TD lasers can be viewed not merely as a matter of energy efficiency, but as being determined by the spatial state of the gain medium and the resulting resonator mode behavior.

When predicting the signal of a TD laser, it is necessary to understand pump absorption, population inversion, gain saturation, and thermal deformation as interrelated energy dynamics in a two-dimensional (2D) spatial domain under steady-state conditions. From an energetic point of view, the steady-state rate equation can provide a sufficient estimation of population inversion and signal power~\cite{RateEq1, MPassPump, MPassPump2, MPassPump3,Bleaching}. However, the resonator mode formed during operation is determined not only by pump-energy extraction but also by wave propagation over the thermally deformed disk surface. Consequently, to improve the agreement between the theoretically predicted signal power and the experimentally observed signal power, a spatial analysis is required to determine how the modified TD affects signal propagation within the cavity and alters mode formation. It becomes particularly important in the high-power regime, where the reduction in signal power cannot be explained solely by an equivalent change in disk curvature. Higher-order aberrations beyond defocus can hinder fundamental-mode build-up, effectively increasing round-trip loss, degrading beam quality, and reducing signal power.

To address these limits, we derive the absorbed pump power per pass and extend the model to 2D transverse intra-cavity field propagation and energy transfer. The model gives a steady-state solution from single-pass to infinite-pass pumping and includes spatially accumulated energy, temperature-dependent cross sections, and nonlinear saturation. In addition, we analyze the factors contributing to the degradation of the transverse mode of the signal through a Zernike-based higher-order aberration analysis of the measured TD. The model was validated experimentally by measuring the residual fraction of pump power from the Yb:YAG thin-disk module. Then, it is applied to the Yb:YAG laser system by combining measured disk optical path difference (OPD) maps with the rate equation to describe the cavity dynamics. 

\section{Materials and methods}
The numerical implementation for analysis consists of two parts: (i) calculating the locally accumulated pump intensity and effective absorption coefficient while including absorption saturation, gain saturation, and the spatial intensity distribution at a given temperature; and (ii) combining intra-cavity field propagation with signal amplification to obtain the cavity mode. 

\subsection{Steady-state absorption in multi-pass pumping}
The spatial behavior of the intra-cavity field within the resonator leads to spatial inhomogeneities in the population inversion. We derive the effective absorption coefficient under non-lasing conditions to account for the spatial inhomogeneity. Saturation is included so that the absorption and gain calculations can be coupled to 2D propagation. 

Saturation arises from changes in the level of population densities: stimulated emission depletes the excited-state while ground-state absorption depletes the ground-state~\cite{Cross1,Cross2}. Because both processes act on the same ion ensemble, any change in the upper-state population that controls signal power also affects pump absorption, even when the transitions occur at different wavelengths. Thermalization of the quantum-defect energy further alters line-widths, cross-sections and population distributions, reinforcing this mutual coupling~\cite{Cross3,Cross4}. The absorption dynamics of Yb:YAG can be adequately described by a two-level system since rapid intra-manifold thermal relaxation instantaneously equilibrates the electronic population within each of the $\mathrm{^2F_{7/2}}$ and $\mathrm{^2F_{5/2}}$ manifolds. Here, excited-state absorption, cross relaxation and energy transfer upconversion are neglected due to the structure of $\mathrm{Yb^{3+}}$~\cite{YbArt1, YbArt2}.

Assuming monochromatic pumping into a medium of thickness $L$ and cross-sectional area $A_{p}$, we define the steady-state effective absorption coefficient $\alpha_\mathrm{eff}$ by equating the local pump-photon flux to the rate of ground-state depletion under steady-state rate equations.  In order to compute the 2D steady‑state inversion prior to lasing, we first derive the non‑lasing effective pump absorption coefficient $(\alpha_\mathrm{eff})$, which maps the multi‑pass pump intensity to the local absorbed pump rate and thus directly determines 2D population inversion ratio. At the steady-state temperature $(T_\mathrm{ss})$, $\alpha_\mathrm{eff}$ is defined using saturation intensity at pump wavelength $(I^\mathrm{(p)}_\mathrm{sat})$. $I^\mathrm{(p)}_\mathrm{eff}$ is the effective pump intensity defined by the spatial overlap of the transmitted pump beams during multi-pass pumping.

\begin{align} \label{eqn:SatAbs}
\alpha_0 = N_0 \sigma_\mathrm{abs}^\mathrm{(p)}(T_\mathrm{ss}),\quad\alpha_\mathrm{eff} = \dfrac{\alpha_0}{1+I^\mathrm{(p)}_\mathrm{eff}/I^\mathrm{(p)}_\mathrm{sat}}
\end{align}

where $N_0 = N_1+N_2$, total ion density with lower laser manifold ($N_1$) and upper laser manifold ($N_2$), $\tau_f$ is the fluorescence lifetime, and $\sigma^\mathrm{(p)}_\mathrm{abs}(T_\mathrm{ss})$ is the absorption cross-section at $T_\mathrm{ss}$. We approximate the manifold to two levels using the quasi-three level rate equation, which is valid for media such as Yb:YAG~\cite{LV3}. Eq. (\ref{eqn:SatAbs}) is called the absorption saturation effect derived from the population inversion equation (See \ref{app1}), and effectively captures the reduction in the absorption rate induced by population inversion. In the steady-state response, the effective absorption $\alpha_\mathrm{eff}$ must follow

\begin{align} \label{eqn:SS}
\begin{split}
\frac{\partial \alpha_\mathrm{eff}}{\partial t} = 0,\quad ( 0\leq\alpha_\mathrm{eff}\leq\alpha_0)
\end{split}
\end{align}

The saturation intensity for the pump and signal wavelength $(p,l)$ with fluorescence lifetime $\tau_f$

\begin{align} \label{eqn:2}
I_\mathrm{sat}^\mathrm{(p,l)} = \frac{h\nu^\mathrm{(p,l)}}{(\sigma^\mathrm{(p,l)}_\mathrm{abs}(T_\mathrm{ss})+\sigma^\mathrm{(p,l)}_\mathrm{em}(T_\mathrm{ss}))\tau_f}
\end{align}

Measured cross-section values include the thermal population of Stark-manifolds and reabsorption/re-emission effects, so individual Stark sublevels do not need to be modeled explicitly. Alternatively, theoretical cross sections can be used~\cite{MPassPump}. For an incident pump power $(P_0)$ from incident pump energy $E_0$, the incident pump intensity is defined as $I_\mathrm{p}= P_0/A_\mathrm{p}$ with  cross-sectional area ($A_\mathrm{p}$). After $n_\mathrm{pass}$ passes, the transmitted pump beams spatially overlap and give the accumulated intensity $I_\mathrm{eff}^\mathrm{(p)}$ as

\begin{align} \label{eqn:OVEq}
\begin{split}
I_\mathrm{eff}^\mathrm{(p)} = \dfrac{dE_0}{A_\mathrm{p}dt} e^{-\alpha_\mathrm{eff}L}\sum_{k=0}^{n_\mathrm{pass}-1}e^{-k\alpha_\mathrm{eff}L}\\
=I_\mathrm{p}e^{-\alpha_\mathrm{eff}L}\frac{1-e^{-n_\mathrm{pass}\alpha_\mathrm{eff}L}}{1-e^{-\alpha_\mathrm{eff}L}}\\
\end{split}
\end{align}
 
Simultaneously, the effective pump intensity ($I^\mathrm{(p)}_\mathrm{eff}$) is defined with spatially overlapped energy in unit time $(dE_\mathrm{OL}/dt)$. The effective pump intensity ($I^\mathrm{(p)}_\mathrm{eff}$) in the gain media depends on the number of passes and the effective local absorption coefficient $\alpha_\mathrm{eff}$ over the multi-pass pump path. To find the $\alpha_\mathrm{eff}$, the equation Eq. (\ref{eqn:SCEq}) must therefore be solved, which is derived by Eqs. (\ref{eqn:OVEq}) and (\ref{eqn:SatAbs}), under steady-state response

\begin{align} \label{eqn:SCEq}
\alpha_\mathrm{eff}=\dfrac{\alpha_0}{1+\dfrac{I_\mathrm{p}}{I_\mathrm{sat}^{\mathrm{(p)}}}\dfrac{e^{-\alpha_\mathrm{eff}L}(1-e^{-n_\mathrm{pass}\alpha_\mathrm{eff}L})}{1-e^{-\alpha_\mathrm{eff}L}}}
\end{align}

It highlights that the effective pump intensity ($I^\mathrm{(p)}_\mathrm{eff}$) and effective absorption coefficient $(\alpha_\mathrm{eff})$ uniquely follow from a steady-state formulation. The $\alpha_\mathrm{eff}$ should not change in a steady-state. Thus, from Eq. (\ref{eqn:SCEq}), $\alpha_\mathrm{eff}$ satisfies
\begin{align} \label{eqn:FP}
\alpha_\mathrm{eff} = f(\alpha_\mathrm{eff},I_\mathrm{p})
\end{align}
Eq. (\ref{eqn:FP}) guarantees that only one finite solution \( \alpha_\mathrm{eff}^*(I_\mathrm{p}) \) exists for each \( I_\mathrm{p}> 0 \) (See the \ref{app1} for the detailed proof).For a prescribed incident pump intensity $I_\mathrm{p}$, the effective absorption coefficient $\alpha_\mathrm{eff}^*$ is obtained, which thus determines effective pump intensity $I_\mathrm{eff}^\mathrm{(p)}$ from Eq. (\ref{eqn:OVEq}). Under the uniform-field approximation, the absorbed pump intensity is obtained by integrating the local absorption rate is

\begin{align}\label{eqn:absP}
\begin{split}
\frac{P_\mathrm{abs}}{A_\mathrm{p}} = -\int_0^L{\frac{dI(z)}{dz}dz} = \int_0^L{\alpha(z)I(z)dz} \\
= I_\mathrm{eff}^\mathrm{(p)}(1-e^{-\alpha_\mathrm{eff}L})
\end{split}
\end{align}

The TD geometry meets $\alpha_\mathrm{eff}L\ll 1$, thus the absorbed power density is then, 
\begin{align}\label{eqn:absP2}
    \frac{P_\mathrm{abs}}{A_\mathrm{p}}\approx\alpha_\mathrm{eff}I^\mathrm{(p)}_\mathrm{eff}L
\end{align}

so that $\alpha_\mathrm{eff}I^\mathrm{(p)}_\mathrm{eff}$ denotes the local absorbed pump power density $(W\cdot m^{-3})$. From Eq. (\ref{eqn:SatAbs}), the absorbed power density is
\begin{align}\label{eqn:Intuitive}
\begin{split}
\mathcal{A}(I^\mathrm{(p)}_\mathrm{eff}) \equiv \alpha_\mathrm{eff}I^\mathrm{(p)}_\mathrm{eff}= \frac{\alpha_0 I^\mathrm{(p)}_\mathrm{eff}}{1+I^\mathrm{(p)}_\mathrm{eff}/I^\mathrm{(p)}_\mathrm{sat}}\\=\alpha_0I_\mathrm{sat}^\mathrm{(p)}\frac{x}{1+x}\quad
\text{where}\quad x\equiv \frac{I^\mathrm{(p)}_\mathrm{eff}}{I^\mathrm{(p)}_\mathrm{sat}}
\end{split}
\end{align}
which increases monotonically with $I^\mathrm{(p)}_\mathrm{eff}$ and tends to the finite limit $\alpha_0 I^\mathrm{(p)}_\mathrm{sat}$ as $x\rightarrow\infty$. Hence, although $\alpha_\mathrm{eff}\rightarrow 0$ for very large $I^\mathrm{(p)}_\mathrm{eff}$, the absorbed power density does not vanish but saturates; the absorbed power per unit area tends to $\alpha_0 I^\mathrm{(p)}_\mathrm{sat}L$ in the uniform field approximation.

\begin{figure}[t]
\centering
\includegraphics[width=\textwidth]{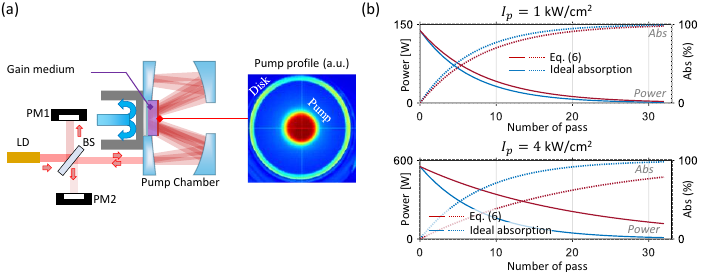}
\caption{\label{fig_1} (a) Experimental configuration for residual pump-power measurement. (b) Typical calculation results of the residual pump power after absorption with (Eq. (\ref{eqn:FP})) and without saturation effects (Ideal absorption) for incident pump intensities ($I_\mathrm{p}$) of 1 $\mathrm{kW/cm^2}$ and 4 $\mathrm{kW/cm^2}$, respectively. Note pump intensity $I_\mathrm{p}$ is defined as the incident pump power divided by the pump beam area. Abbreviations are as follows; LD: laser diode, PM: power-meter and BS: beam splitter.}. 
\end{figure}

In the low pump intensity limit $(I_\mathrm{eff}^\mathrm{(p)}\ll I_\mathrm{sat}^\mathrm{(p)})$, Eq. (\ref{eqn:OVEq}) admits a Taylor expansion whose leading term yields the familiar linear scaling from Eq. (\ref{eqn:OVEq}),  reduced to the linear approximation 
\begin{align}\label{eqn:LowAbs}
I^\mathrm{(p)}_\mathrm{eff} \approx \frac{P_0}{A_{p}}n_\mathrm{pass}(1-\frac{n_\mathrm{pass}+1}{2}\alpha_0L)
\end{align}
which validates the expressions employed in this analysis of multi-pass pumping. In the limit $I^\mathrm{(p)}_\mathrm{eff}\gg I^\mathrm{(p)}_\mathrm{sat}$ (i.e., when the incident pump power is much larger than the saturation intensity or, equivalently, when the absorption cross section is very small), Eq. (\ref{eqn:OVEq}) reduces to

\begin{align}\label{eqn:HighAbs}
I^\mathrm{(p)}_\mathrm{eff} \approx \dfrac{P_0}{A_\mathrm{p}}\times n_\mathrm{pass}
\end{align}

Since achieving the condition of $(I^\mathrm{(p)}_\mathrm{eff}\gg I^\mathrm{(p)}_\mathrm{sat})$ is unattainable for the gain media of interest and practically undesirable, it is essential to derive a solution $\alpha_\mathrm{eff}^*(I_\mathrm{p})$. If we assume $n_\mathrm{pass}\rightarrow\infty$, the effective pump intensity in the medium can be defined as Eq. (\ref{eqn:8}). Assuming a large number of passes, it is proper to derive the solution $\alpha_\mathrm{eff}^*(I_\mathrm{p})$ through Eqs. (\ref{eqn:8}) and (\ref{eqn:SatAbs}). 
\begin{align}\label{eqn:8}
\begin{split}
    \lim_{n_\mathrm{pass}\rightarrow\infty}I^\mathrm{(p)}_\mathrm{eff} =\dfrac{I_\mathrm{p}e^{-\alpha_\mathrm{eff}L}}{1-e^{-\alpha_\mathrm{eff}L}}
\end{split}
\end{align}

Fig. \ref{fig_1}(a) shows the experimental setup used to measure the residual pump fraction in the 32 pass Yb:YAG thin-disk module. As seen in Fig. \ref{fig_1}(b), the difference in effective pump intensity between relatively low and high incident intensities is clearly observed. The derived $\alpha_\mathrm{eff}^*(I_\mathrm{p})$ demonstrates the effectiveness in the high-intensity pumping regime. The reconstructed absorption map is shown in Fig. \ref{fig_2}(d). Further discussion and comparison of the results will be presented in section \ref{RnD}.

\subsection{Signal photon evolution}
\begin{figure}[t]
\centering
\includegraphics[width=\textwidth]{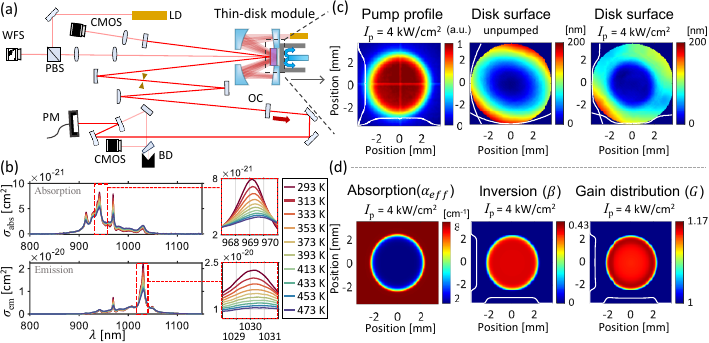}
\caption{\label{fig_2} (a) Experimental setup: simultaneous measurement of signal power, beam profile, and disk wavefront during pumping. (b) Temperature-dependent absorption and emission cross-section of Yb:YAG. (c) Typical example of measured 2D distribution: pump profile, disk surface when unpumped and $I_\mathrm{p}=$ 4 $\mathrm{kW/cm^2}$ pumped. (d) Typical example of simulated 2D distributions: effective absorption coefficient, initial population inversion ratio ($\beta$) prior to lasing, and initial single-pass gain ($G_D$) prior to lasing at the pump intensity of 4 $\mathrm{kW/cm^2}$. Abbreviations as follows; WFS: wavefront sensor, LD: laser diode, PM: power-meter, BD: beam dump, PBS: polarized beam splitter and OC: output coupler.} 
\end{figure}

Pump photon flux $(\Phi_\mathrm{p})$ is derived from effective pump intensity $\Phi_{p} ={I^\mathrm{(p)}_\mathrm{eff}}/{h\nu^\mathrm{(p)}}$. We avoid updating the absorption coefficient via the local population inversion at every propagation step by precomputing a 2D effective pump absorption $\alpha_\mathrm{eff}$ from the multi-pass pump distribution and using it as the fixed input for inversion initialization, which substantially reduces computational cost and improves loop efficiency in 2D cavity propagation. Pump photon flux, reflecting pump saturation effect and estimated Gaussian order shown in Fig.\ref{fig_2}(c), replaces the usual single-pass intensity in the quasi-three level rate equation defined by population inversion ratio $\beta=N_2/N_0$. The experimental setup for Yb:YAG TD laser is shown in Fig. \ref{fig_2}(a). For quasi-three level media such as Yb:YAG, where the lower laser level lies near the ground-state and rapidly decays to it, thermal excitation partially repopulates the lower level, allowing a quasi-three level approximation~\cite{LV3}.

\begin{align}\label{eqn:beta}
\begin{split}
\dfrac{d\beta}{dt} = -\gamma_{12}\beta+\big(\sigma_\mathrm{abs}^\mathrm{(p)}(1-\beta)-\sigma_\mathrm{em}^\mathrm{(p)}\beta\big)\Phi_\mathrm{p}\\+\big(\sigma_\mathrm{abs}^\mathrm{(l)}(1-\beta)-\sigma_\mathrm{em}^\mathrm{(l)}\beta\big)\Phi_\mathrm{l}
\end{split}
\end{align}

where $\gamma_{12}=1/\tau_f$ is the decay rate and $\Phi_\mathrm{l}$ is signal photon flux in unit area that $\Phi_\mathrm{l} ={I_\mathrm{l}}/{h\nu^\mathrm{(l)}}$ with signal intensity ($I_\mathrm{l}$). Eq. (\ref{eqn:beta}) ensures that gain saturation, temperature dependent cross-sections, and spatial pump beam distribution are all captured. The absorption and emission cross-sections used in this equation are measured values corresponding to $T_\mathrm{ss}$, as shown in Fig. \ref{fig_2}(b). The $I^\mathrm{(p)}_\mathrm{eff}$ is utilized for deriving pump photon flux, then small signal gain $g_0$ is derived via $\beta$ 
\begin{align}\label{eqn:10}
\begin{split}
g_0(\beta) = N_0(\sigma_\mathrm{em}^\mathrm{(l)}\beta-\sigma_\mathrm{abs}^\mathrm{(l)}(1-\beta))\\
G_D(\beta)=e^{(g_\mathrm{eff}L)}
\end{split}
\end{align}

The intra-cavity signal photon flux $(\Phi_\mathrm{l})$ can be derived with an effective small signal gain $g_\mathrm{eff} = {g_0}/{(1+I_\mathrm{l}/I^\mathrm{(l)}_\mathrm{sat})}$ and then used to derive the single-pass gain $G_D$. In the numerical implementation, pump-side absorption initialization and signal-side gain saturation are treated in a split, quasi-static separated manner that do not interact at leading order. The population inversion $\beta$, derived in the non-lasing limit, captures pump saturation, while the effective gain $g_{\mathrm{eff}}$ accounts for signal saturation without double counting, since these operate on distinct physical processes and rate-equation terms. This decomposition helps identify the spatial region where net gain supports lasing lasing occurs and eliminates the need for iterative absorption equilibration and enables rapid convergence to signal photon flux and spatially-resolved gain. 

This quasi-static treatment is essential for efficient 2D spatial modeling. The first reason is due to spatial inhomogeneities in the region where the pump beam and stimulated emission interact. Second, a full iterative solution of coupled population-field dynamics at each spatial point would impose prohibitive computational cost while delivering only marginal refinement in macroscopic output prediction. Instead, we compute the pump-induced steady-state inversion profile once from the input pump distribution, then iterate the signal field to convergence. The resulting intra-cavity photon flux and spatial gain map thus captures the essential physics of mode-dependent inversion inhomogeneity in regions of single-mode dominance, while achieving rapid convergence. Validation from low to high-power regimes confirms that this efficient approach provides accurate prediction of lasing performance while retaining the dominant physics needed to reproduce the measured output characteristics.

\subsection{Roundtrip propagation}
The derived signal photon flux is utilized to determine the precise distribution of the steady-state electric field $u(x,y)$ inside the cavity, $x$ and $y$ being the transverse coordinates and $z$ the direction along the cavity. The intra-cavity electric field arises from repeated amplification, diffractive propagation, and superposition of the emission source~\cite{Hodgson,Diff1}. To calculate it, we therefore apply operators related to these 3 effects to an initial distribution of the electric field, $u_{in}(x, y)$. We then iterate until we reach convergence of the solution. 

The seed term is chosen as $u_\mathrm{in}(x,y) = u_\mathrm{SE}(x,y)$, a spatially localized spontaneous emission sourcefor a given pump intensity~\cite{ASE_Loss}. For the propagation, the length of the cavity is divided into $N$ segments of distance $d_j(j\in[1,2,...,N])$, and the combined propagation operator $\mathcal{P}_j$ over the segment includes the Rayleigh–Sommerfeld diffraction $\mathcal{R}_j$ and the complex phase $\mathcal{M}_j$ of the piece of optics present in the given segment.

\begin{align}\label{PropOp}
\mathcal{P}_j\{u(x,y)\}\equiv \mathcal{M}_j\circ\mathcal{R}_j\{u(x,y)\} 
\end{align}

 The disk-induced phase difference is defined in terms of the total surface profile $S_{\rm D,tot}(x,y)$ and the vacuum wavenumber $k_0$ as

\begin{align}\label{surf_phase}
\begin{split}
    \mathcal{M}\{S_\mathrm{D,tot}\} = e^{-i\cdot 2 k_0\cdot S_\mathrm{D,tot}}=e^{-i\Psi_\mathrm{D}}
\end{split}
\end{align}

where $\Psi_\mathrm{D}$ is the OPD of the TD surface, capturing the effects of mechanical, thermal-expansion, and air convection, which are thus naturally accounted for the diffraction calculation~\cite{Diff1,Convect1}. The surface deformation has been taken from experimental measurement at different pump intensities, considering a factor of 2 due to the reflection configuration of our measurement. The measured disk surface has been utilized for the disk-induced phase(see Fig. \ref{fig_2}(c)). Composing the segment operators gives the roundtrip operator $\mathcal{T}$

\begin{align}\label{RT}
\mathcal{T}\{u(x,y)\} \equiv \mathcal{P}_N\circ\mathcal{P}_{N-1}\circ...\circ\mathcal{P}_2\circ\mathcal{P}_1\{u(x,y)\}
\end{align}

\begin{figure}[t]
\centering
\includegraphics[width=\textwidth]{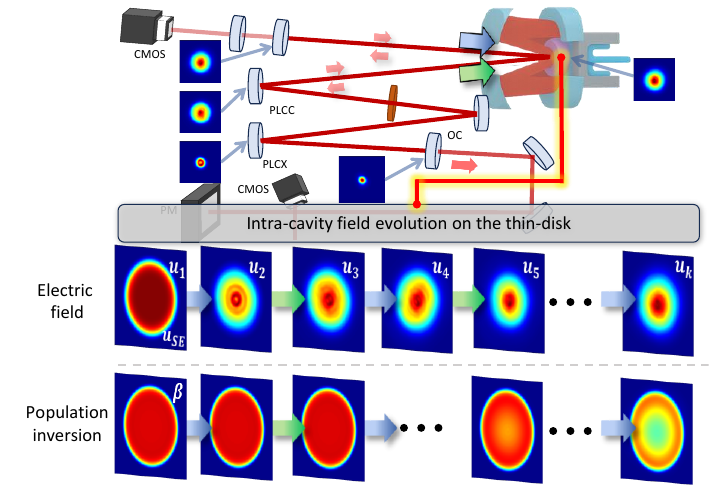}
\caption{\label{fig_3} Illustration of simulation for the evolution of the intra-cavity signal field with population inversion. The electric field propagates and amplifies within the resonator, leading to transverse-mode formation.}
\end{figure}

After the $k$-th disk encounter, the intra-cavity field is updated as 
\begin{align}\label{wave_prop}
\begin{split}
    u_{k+1}(x,y) = u_\mathrm{SE}(x,y) + \sqrt{(1-\alpha_l)\cdot G_\mathrm{D}(x,y)} e^{[-i\Psi_\mathrm{D}(x,y)]}\mathcal{T}\{u_{k}(x,y)\}
        \\(k = 1,2,...)
\end{split}
\end{align}
where $\alpha_l$ is the total intra-cavity loss, including the transmission of the output coupler as $5\%$ and the roundtrip cavity losses of 1$\%$. After each round trip, the intra-cavity intensity is evaluated as $I_\mathrm{l} = |u_k|^2$. The system is considered to have reached a steady state when ($(|u_{k-1}|^2-|u_{k}|^2)/|u_{k}|^2< \epsilon$. Note that $u_k(x,y)$ represents the intra-cavity field prior to output coupling. The intra-cavity field $u_k(x,y)$ evolves sequentially, as shown in Fig. \ref{fig_3}.

\subsection{Experimental configuration}
We first validated the approach by measuring the residual pump power (Fig. \ref{fig_1}(a)) and subsequently confirmed it in the single-mode laser by comparing predicted and measured 2D signal fields(Fig. \ref{fig_2}(a)). A $7.9\,\mathrm{at.\%}$, 188 $\mu m$-thick Yb:YAG disk was used in the experimental setup, with the incident pump intensity $I_\mathrm{p}$ was varied from 0 to 4 $\mathrm{kW/cm^2}$. The measured pump distribution was fitted with a sixth order super-Gaussian and used as the pump input. During the laser operation, the TD wavefront was measured in situ, and the deformed surface of TD was recorded 40 times, with the same procedures as Refs~\cite{Muzik_2015,Sikocinski_2016}. The averaged surface was interpolated into the fine grid, then used to define the OPD in the cavity calculation. The tilt term was removed using Zernike decomposition. Beam diameter was evaluated according to the ISO 11146 standard~\cite{ISO11146} and the $\mathrm{M^2}$ parameter was obtained by deriving the covariance matrix from the electric field~\cite{M2}.

\subsection{Numerical implementation}

The numerical calculations of beam propagation, pump absorption and gain evolution were carried out by MATLAB 2023a. The grid size of the 2D field was chosen based on the minimum distance between optical elements. Measured temperature-dependent absorption and emission cross-sections \cite{MeasYBYAG} were used to calculate the initial 2D population inversion and single-pass gain distribution, as shown in Fig. \ref{fig_2}(d). The disk temperature field was calculated using a Matlab-based energetics model combined with steady-state finite-element thermal analysis in COMSOL by considering pump intensity distribution with complex structure of the disk (diamond heatsink, disk bonding layer, coatings and active medium), following the procedure described in Refs.~\cite{Slezak2013,Slezak2014}. The derived surface temperature of the TD was validated against measured surface temperatures(Teledyne FLIR A65). The signal beam was observed 50 cm from the output coupler, and the simulation used the same observation plane. The parameters utilized for the numerical calculation are listed in the Table \ref{Tab_Param} (\ref{app_Param}).

In addition, to interpret the measured disk deformation, the reflected OPD within the pump aperture was expanded into Zernike polynomials, $W(\rho,\theta)=\sum_j C_j Z_j(\rho,\theta)$. The coefficient magnitudes were compared as a function of pump intensity to identify the dominant aberration terms beyond defocus. The defocus coefficient was also converted into an equivalent radius of curvature (eROC). A defocus-only cavity simulation based on this eROC was then compared with the measured-OPD simulation and the experimental data.

\begin{figure}[t]
\centering
\includegraphics[width=\textwidth]{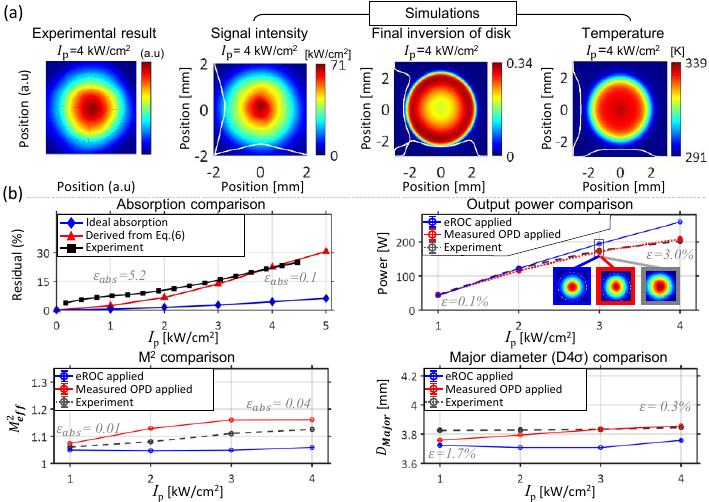}
\caption{\label{fig_4} (a) Typical example of the signal beam of experiments, simulated signal beam profile, the population inversion ratio of the disk in the steady state and simulated temperature profile. (b) Summary of results: comparison between experiment and simulation, where a equivalent radius of curvature (eROC) obtained from the Zernike decomposition. Residual fraction of pump power, signal power, major beam diameter($D_\mathrm{maj}$), and beam quality factor ($\mathrm{M^2}$) are compared. Note, $\varepsilon_{abs}$ is absolute errors between the experiment and simulation result. $\varepsilon$ is relative errors between the experiment and simulation results. Note the simulation result of residual fraction of pump power was derived from Eq. (\ref{eqn:FP}).}
\end{figure}

\clearpage
\section{Results and discussion}
\label{RnD}
The proposed model was evaluated in two steps. First, the non-lasing pump-initialization model was validated by comparing the predicted residual pump fraction with the measurement. Second, the initialized 2D population inversion and gain maps were used in cavity propagation to compare the eROC-only and measured-OPD cases against the laser experiment. As shown in Fig. \ref{fig_4}(a), the numerical model enables the evaluation of the steady-state field, phase, and population inversion of the disk. In each case, the signal converged after an average of 140 iterations. The maximum absolute error in the residual pump rate is 5.2$\%$ at incident pump intensity $I_\mathrm{p} = 1\,\mathrm{kW/cm^2}$. However, the model incorporating the measured OPD yields markedly improved agreement with the experiment (Fig. \ref{fig_4}(b)). Quantitatively, the numerical model reproduces signal power within 3.0$\%$, beam diameters within 1.7$\%$, and $\mathrm{M^2}$ within 0.05 of the measured values.

\begin{figure}[!b]
\centering
\includegraphics[width=\textwidth]{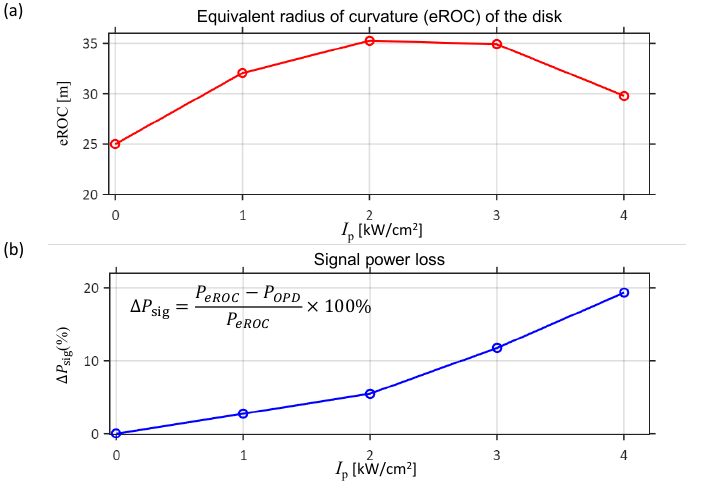}
\caption{\label{fig_5} Signal power reduction beyond the equivalent-curvature approximation (a) Change of eROC of TD according to the incident pump intensity. (b) Signal power difference between the results using eROC and measured OPD.}
\end{figure}

The results obtained by applying the measured OPD from the perspectives of signal power, beam diameter, and beam quality show good agreement with the experiment. The signal power simulated using eROC represents the theoretically derivable signal power. When applying the eROC, the signal beam diameter results do not show the same tendency as the signal power loss(Fig. \ref{fig_5}(a)). Nevertheless, the signal power derived using measured OPD decreases by up to 19.3$\%$ as the incident pump intensity increases, as shown in Fig. \ref{fig_5}(b). When only eROC is applied, the gain distribution serves as a soft aperture for the resonant field. In contrast, the measured disk OPD introduces higher-order phase aberrations that hinders the mode formation. Their combined action reduces the overlap with the supported fundamental cavity mode and increases the effective round-trip loss.

\begin{figure}[!t]
\centering
\includegraphics[width=\textwidth]{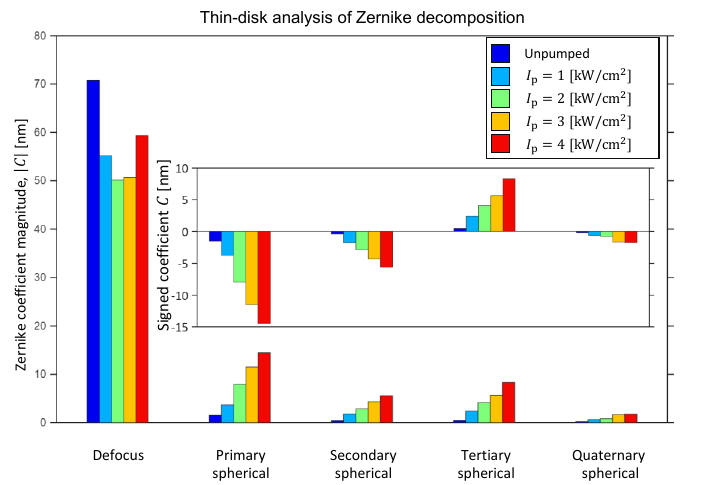}
\caption{\label{fig_6} Zernike decomposition of measured OPD. The figure shows the representative Zernike coefficient magnitudes ($|C|$) trends based on incident pump intensity ($I_\mathrm{p}$) .}
\end{figure}

This trend is further supported by Fig. \ref{fig_6}. As the incident pump power increases, the TD surface becomes more deformed. In the TD we used, low-to-higher spherical aberration tended to increase significantly as the pump intensity increased, as shown in Fig. \ref{fig_6}. The trend is consistent with increasing eROC-to-OPD signal-power discrepancy, shown in Fig. \ref{fig_5}(b). The other higher-order Zernike coefficients did not show a clear trend (See Fig. \ref{fig_C1} and \ref{fig_C2} in \ref{app3} for the detailed result). Therefore, mitigation should focus on flattening the radial heat-load distribution, improving the thermal and mechanical symmetry of the TD assembly, reducing heat generation, and, if necessary, compensating the resonator against residual spherical aberrations.

\begin{figure}[!htbp]
\centering
\includegraphics[width=\textwidth]{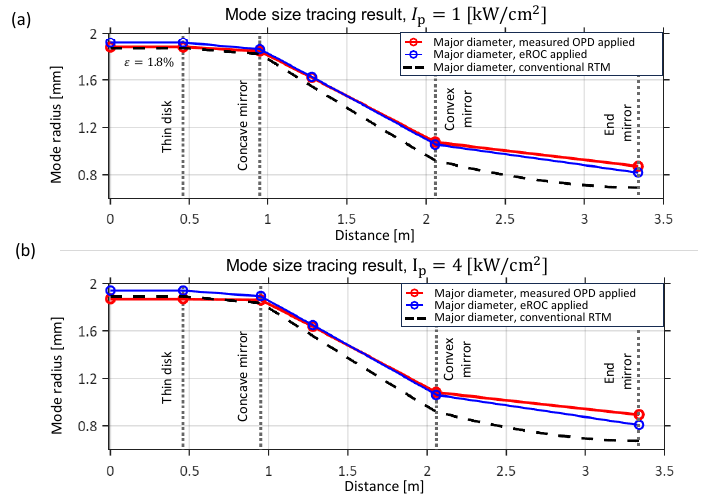}
\caption{\label{fig_7} Intra-cavity mode-size deviation from eROC and ray transfer matrix(RTM) prediction with incident pump intensity (a) 1 $\mathrm{kW/cm^2}$ and (b) 4 $\mathrm{kW/cm^2}$. The black dot line represents the size of the eigenmode from conventional ray transfer matrix (RTM) using the eROC, the red and blue lines represent the numerical calculation result of the major diameter of mode using measured OPD and eROC, respectively.}
\end{figure}

The discrepancy between eROC-only and measured-OPD models is not explained by a simple scalar curvature correction alone; it also appears as a deviation of the circulating mode from the paraxial eigenmode. Fig. \ref{fig_7}(a) and (b) present the results of analyzing the mode diameter inside the cavity using the conventional ray transfer matrix (RTM) with the eROC and the numerical model results with the eROC and the measured OPD for incident pump intensities of 1 $\mathrm{kW/cm^2}$ and 4 $\mathrm{kW/cm^2}$, respectively. The influence of aberrations can be indirectly inferred by comparing the results obtained using the eROC and measured OPD. As the incident pump intensity increases, the mode size on TD with measured OPD decreases by up to approximately 3.6$\%$ compared to the eROC.

For more accurate predictions, the pump-photon flux in laser operation cannot be reliably estimated from a purely static model because absorption and stimulated emission are dynamically coupled through the population inversion. It is necessary to include transient field updates, amplified spontaneous emission (ASE) and spectrally-temperature-resolved populations for improving fidelity. However, it inevitably causes substantial computational cost. Moreover, single-mode operation typically involves a pumped area larger than the signal, producing population inversion inhomogeneity in the spatial domain. Despite these complexities, the present model remains useful as a practical predictive tool for output characteristics under the monochromatic pump and signal assumption.

\section{Conclusion}
We have developed and validated a steady-state analysis framework for high-power thin-disk lasers by combining non-lasing multi-pass pump initialization with two-dimensional cavity propagation based on measured disk OPD. The resulting model reproduces the experimentally observed absorption, signal power, beam diameter, and $\mathrm{M^2}$. To isolate the role of disk deformation, a defocus-only curvature model was used as a reference representing the performance attainable when the measured higher-order aberrations are absent. The comparison between this reference case and the measured-OPD calculation, supported by Zernike decomposition of the disk surface, shows that higher-order aberrations beyond defocus reduce the overlap with the supported fundamental cavity mode and contribute to the signal-power reduction at high pump intensity. These results indicate that power scaling in high-power single-mode thin-disk lasers cannot be assessed from pump-energy extraction or equivalent curvature alone, but requires aberration-resolved cavity analysis based on the measured optical state of the disk.

\section{Funding}
This research was co-funded by the European Union (MERIT - Grant Agreement No. 101081195) and European Union and the state budget of the Czech Republic under the project LasApp CZ.02.01.01/00/22$\_$008/0004573.

\section{Disclosures}
The authors declare no conflicts of interest.

\section{Data Availability Statement}
The data that support the findings of this study are openly available in Zenodo 10.5281/zenodo.17853530

\appendix
\section{Derivation of effective absorption coefficient from quasi-three-level rate equation}
\label{app1}

In this appendix, the effective pump absorption coefficient is derived in the non-lasing limit, i.e., before the intra-cavity signal field becomes large enough to modify the inversion through stimulated emission. The resulting relation is not intended to represent the time evolution of the transient lasing process. Instead, it provides a steady-state pump-side closure that is used to initialize the spatial inversion profile prior to cavity-field iteration. The population inversion ratio in a quasi-three-level system is governed by the rate equation:
\begin{equation}
\frac{d\beta}{dt} = -\gamma_{12} \beta + \left[
  (\sigma_{\mathrm{abs}}^p(1-\beta) - \sigma_{\mathrm{em}}^p\beta)\Phi_\mathrm{p} 
  + (\sigma_{\mathrm{abs}}^l(1-\beta) - \sigma_{\mathrm{em}}^l\beta)\Phi_\mathrm{l}
\right]
\label{eqn:rateEq}
\end{equation}
where the normalized population inversion is defined as:
\begin{align}
\beta &= \frac{N_2}{N_0} \quad \mathrm{(population\ inversion\ ratio)} \\
\Phi_\mathrm{p} &= \frac{I_\mathrm{eff}^\mathrm{(p)}}{h\nu^\mathrm{(p)}} \quad \mathrm{(pump\ photon\ flux)} \\
\Phi_\mathrm{l} &= \frac{I_\mathrm{l}}{h\nu^\mathrm{(l)}} \quad \mathrm{(signal\ photon\ flux)} \\
\gamma_{12} &= \frac{1}{\tau_{f}} \quad \mathrm{(spontaneous\ decay\ rate)}
\end{align}

In the non-lasing limit ($\Phi_\mathrm{l}=0$), the steady-state rate equation reduces to a pump only form.
\begin{equation}
0 = -\gamma_{12}\beta + (\sigma_{\mathrm{abs}}(1-\beta) - \sigma_{\mathrm{em}}\beta)\Phi
\label{eqn:steadyState}
\end{equation}

Saturation occurs when the stimulated transition rate equals the decay rate:
\begin{equation}
(\sigma_{\mathrm{abs}} + \sigma_{\mathrm{em}})\Phi_{\mathrm{sat}} = \gamma_{12}
\label{eqn:saturationCondition}
\end{equation}

Converting to intensity, the saturation intensity is defined as:
\begin{equation}
I_{\mathrm{sat}} = h\nu \Phi_{\mathrm{sat}} = \frac{h\nu\gamma_{12}}{\sigma_{\mathrm{abs}} + \sigma_{\mathrm{em}}}
\label{eqn:satInt}
\end{equation}

Consequently, the effective absorption coefficient exhibits saturable behavior:
\begin{equation}
\alpha_{\mathrm{eff}} = \frac{\alpha_0}{1 + \frac{I^\mathrm{(p)}_\mathrm{eff}}{I^\mathrm{(p)}_{\mathrm{sat}}}}
\label{eqn:satAbs}
\end{equation}

For multi-pass pumping configurations, the spatially overlapped transmitted energy is:
\begin{equation}
I^\mathrm{(p)}_{\mathrm{eff}} = I_\mathrm{p} e^{-\alpha_{\mathrm{eff}}L} \sum_{k=0}^{n_{\mathrm{pass}}-1} e^{-k\alpha_{\mathrm{eff}}L}
= I_\mathrm{p} \cdot \frac{e^{-\alpha_{\mathrm{eff}}L}(1 - e^{-n_{\mathrm{pass}}\alpha_{\mathrm{eff}}L})}{(1 - e^{-\alpha_{\mathrm{eff}}L})}
\label{eqn:overlapIntensity}
\end{equation}

The spatially overlapped intensity ($I^\mathrm{(p)}_{\mathrm{eff}}$)is related to the incident pump intensity ($I_\mathrm{p}$). Substituting Eq.~\eqref{eqn:overlapIntensity} into Eq.~\eqref{eqn:satAbs}, the effective absorption 
coefficient is implicitly defined by:
\begin{equation}
\alpha_{\mathrm{eff}} = f(\alpha_{\mathrm{eff}}, I_\mathrm{p}) = 
\dfrac{\alpha_0}{1 + \dfrac{I_\mathrm{p}}{I_\mathrm{sat}^\mathrm{(p)}} \cdot \dfrac{e^{-\alpha_{\mathrm{eff}}L}(1 - e^{-n_{\mathrm{pass}}\alpha_{\mathrm{eff}}L})}{(1 - e^{-\alpha_{\mathrm{eff}}L})}}
\label{eqn:selfConsistent}
\end{equation}

where the implicit steady-state equation is:
\begin{equation}
\alpha_{\mathrm{eff}}^* = f(\alpha_{\mathrm{eff}}^*, I_\mathrm{p})
\label{eqn:fixedPoint}
\end{equation}
with the domain constraints $(\alpha_0 > 0)$, $(0<\alpha_\mathrm{eff}<\alpha_0), (n_{\mathrm{pass}} \geq 1)$, and $(I_\mathrm{p} > 0)$. $x=\alpha_{\rm eff}L$, $C=\alpha_0L$, $B=I_p/I_{\rm sat}^{(p)}$, and

\begin{equation}
S(x)=\sum_{j=1}^{n_{\rm pass}}e^{-jx},
\end{equation}
Eq.~(\ref{eqn:selfConsistent}) can be written as

\begin{equation}
x=\frac{C}{1+B S(x)}.
\end{equation}
Equivalently,
\begin{equation}
B=T(x), \qquad
T(x)=\frac{C-x}{xS(x)} .
\end{equation}

On the admissible interval $0<x<C$, $T(x)$ is positive and continuous,
with $T(x)\rightarrow\infty$ as $x\rightarrow0^+$ and
$T(x)\rightarrow0$ as $x\rightarrow C^-$. To establish uniqueness, we
differentiate $T(x)$ logarithmically:
\begin{equation}
    \frac{T'(x)}{T(x)}
=-\frac{1}{C-x}-\frac{1}{x}-\frac{S'(x)}{S(x)}
= \mu(x)-\frac{C}{x(C-x)},
\end{equation}

where
\begin{equation}
    \mu(x)= \frac{\sum_{j=1}^{n_{\rm pass}}j e^{-jx}}
{\sum_{j=1}^{n_{\rm pass}}e^{-jx}}
\end{equation}

Since $\mu(x)<(n_{\rm pass}+1)/2$ and $C/[x(C-x)]\ge 4/C$, $T'(x)<0$ is guaranteed when
\begin{equation}
    C<\frac{8}{n_{\rm pass}+1}.    
\end{equation}

This condition is satisfied for the present thin-disk configuration.
Therefore, $T(x)$ maps $(0,C)$ monotonically from $\infty$ to $0$.
Consequently, for every prescribed $B>0$, and hence for every
$I_p>0$, Eq.~(\ref{eqn:selfConsistent}) has exactly one solution
$x\in(0,C)$, corresponding to a unique
$\alpha_{\rm eff}=x/L\in(0,\alpha_0)$

\clearpage

\section{Numerical calculation parameters}
\label{app_Param}
\begin{table}[!htbp]
\scriptsize
\centering
\caption{Simulation parameters}
\label{Tab_Param}
\begin{tabular}{@{}lcccc@{}}
\hline
Symbol & unit & value & description & Notes \\
\hline
$\sigma_\mathrm{abs}^\mathrm{(p)}$ & $\mathrm{cm^2}$ & \makecell{7.54$\times 10^{-21}$ at 27$^\circ$C\\5.11$\times 10^{-21}$ at 100$^\circ$C} & Absorption cross-section at $\lambda_\mathrm{p}$ & shown in Fig. \ref{fig_2}(b) \\
$\sigma_\mathrm{abs}^\mathrm{(l)}$ & $\mathrm{cm^2}$ &\makecell{1.25$\times 10^{-21}$ at 27$^\circ$C\\1.53$\times 10^{-21}$ at 100$^\circ$C} & Absorption cross-section at $\lambda_\mathrm{l}$ & shown in Fig. \ref{fig_2}(b) \\
$\sigma_\mathrm{em}^\mathrm{(p)}$ & $\mathrm{cm^2}$ &\makecell{6.88$\times 10^{-21}$ at 27$^\circ$C\\4.66$\times 10^{-21}$ at 100$^\circ$C} & Emission cross-section at $\lambda_\mathrm{p}$ & shown in Fig. \ref{fig_2}(b) \\
$\sigma_\mathrm{em}^\mathrm{(l)}$ & $\mathrm{cm^2}$ &\makecell{2.15$\times 10^{-20}$ at 27$^\circ$C\\1.49$\times 10^{-20}$ at 100$^\circ$C} & Emission cross-section at $\lambda_\mathrm{l}$ & shown in Fig. \ref{fig_2}(b) \\
$L$ & $\mu$m & 188 & TD thickness &- \\
$N_0$ & $\mathrm{cm^{-3}}$ & 1.094$\times 10^{21}$ &  Total ion density & - \\
$\tau_f$ & $\mathrm{ms}$ & 0.95 & Fluorescence lifetime & - \\
$\gamma_{12}$ & kHz & 1.052 & Decay rate & $\gamma_{12} = 1/\tau_f$ \\
$n_\mathrm{pass}$ & - & 32 & Number of passes for pumping & - \\
$T_\mathrm{ss}$ & K & \makecell{ 291 at unpumped \\ 339 at 4 $\mathrm{kW/cm^2}$} & TD temperature & varies with pump intensity\\
$\lambda_\mathrm{p}$ & nm & 969 & Pump wavelength & - \\
$\lambda_\mathrm{l}$ & nm & 1030 & Signal wavelength & - \\
$\alpha_l$ & $\%$ & 6 & Total intra-cavity loss & include OC transmission ($5\%$)\\
$R_p$ & mm & 2.05 & Pump radius, HWHM & -\\
$n_g$ & - & 6 & Gaussian order of pump beam & - \\
$N_\mathrm{grid}$ & - & 4096 & Grid size &  \makecell{ Minimum distance considered \\ shown in Fig. \ref{fig_B}} \\
$D_x$ & mm & 25.4 & Window size, x-axis & shown in Fig. \ref{fig_B} \\
$D_y$ & mm & 25.4 & Window size, y-axis & shown in Fig. \ref{fig_B} \\
$p_x$ & $\mu$m & 6.2 & Pixel pitch, x-axis & shown in Fig. \ref{fig_B} \\
$p_y$ & $\mu$m & 6.2 & Pixel pitch, y-axis & shown in Fig. \ref{fig_B} \\
$R_{zern}$ & mm & 2.05 & \makecell{Aperture radius for \\ Zernike decomposition}  & same as pump radius \\

\hline
\multicolumn{5}{c}{\textit{Abbreviation follows:} OC: output coupler, TD: thin-disk, HWHM: half-width-half-maximum }

\end{tabular}

\end{table}
\begin{figure}[!htbp]
\renewcommand{\thefigure}{B\arabic{figure}}
\setcounter{figure}{0}
\centering
\includegraphics[width=\textwidth]{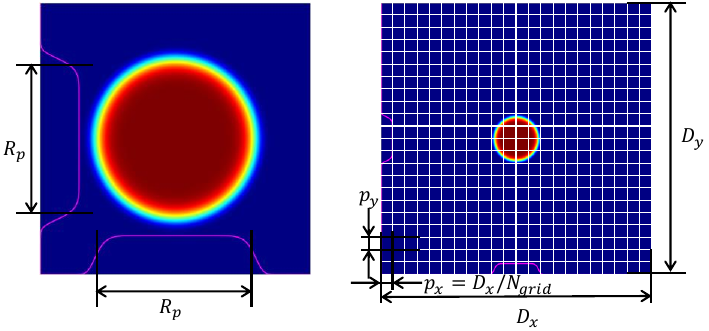}
\caption{\label{fig_B} Parameter description for beam}
\end{figure}
\clearpage

\section{Zernike decomposition result}
\label{app3}
\begin{figure}[!htbp]
\renewcommand{\thefigure}{C\arabic{figure}}
\centering
\includegraphics[width=\textwidth]{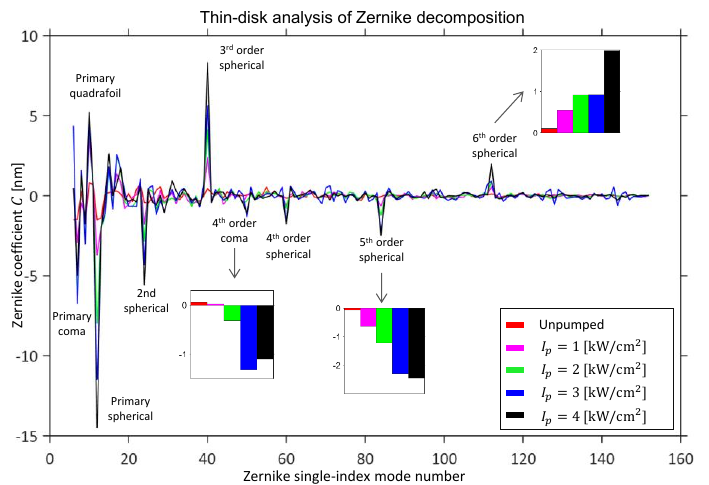}
\caption{\label{fig_C2} Zernike decomposition result for higher-order aberrations. The piston, tilt, defocus, and primary astigmatism terms are excluded}
\end{figure}

We performed a Zernike decomposition on the measured surface. Zernike coefficient for the low-order aberrations is shown in Fig. \ref{fig_C1}. For a clear comparison, the piston, tilt, defocus, and primary astigmatism terms (p=0–5) were excluded from the plot in Fig. \ref{fig_C2}. Single-index Zernike mode $p$ can be converted into Azimuthal order $m$ and Radial order $n$~\cite{Zern}.

\begin{figure}[!htbp]
\renewcommand{\thefigure}{C\arabic{figure}}
\setcounter{figure}{0}
\centering
\includegraphics[width=\textwidth]{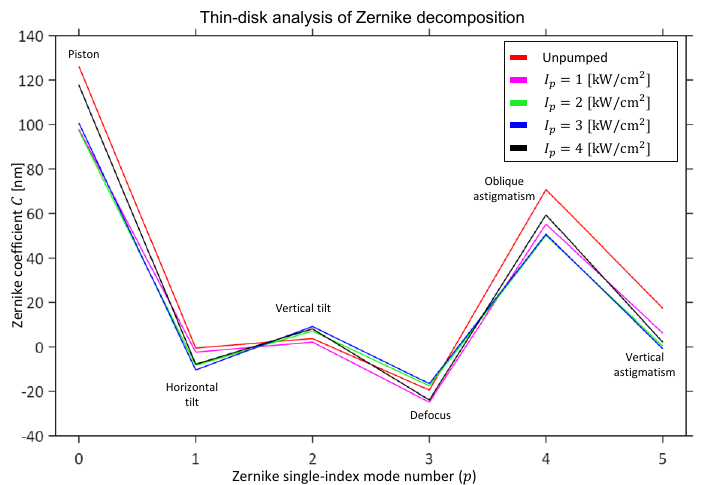}
\caption{\label{fig_C1} Zernike decomposition result for lower-order aberrations. The piston, tilt, defocus, and primary astigmatism terms are included}
\end{figure}
\clearpage

\clearpage



\bibliographystyle{elsarticle-num} 
\bibliography{reference}

\end{document}